\documentclass{iaus}
\usepackage{graphicx}

\def\spose#1{\hbox to 0pt{#1\hss}}
\def\simlt{\mathrel{\spose{\lower 3pt\hbox{$\mathchar"218$}}
     \raise 2.0pt\hbox{$\mathchar"13C$}}}
\def\simgt{\mathrel{\spose{\lower 3pt\hbox{$\mathchar"218$}}
     \raise 2.0pt\hbox{$\mathchar"13E$}}}

\title[Cosmological Disk Simulations on a Mesh]
{Hydrodynamical Adaptive Mesh Refinement Simulations of Disk Galaxies}

\author[B.~K. Gibson {\it et~al.}]
{Brad K. Gibson$^1$\thanks{\tt http://www.uclan.ac.uk/$\sim$bkg/},
St\'ephanie Courty$^1$, Patricia S\'anchez-Bl\'azquez$^1$,
Romain Teyssier$^2$, Elisa L. House$^1$,\\ 
Chris B. Brook$^3$, \and Daisuke Kawata$^4$}

\affiliation{$^1$Centre for Astrophysics, University of Central Lancashire, 
Preston, PR1~2HE, UK \\
email: {\tt bkgibson@uclan.ac.uk} \\[\affilskip]
$^2$Service d'Astrophysique, CEA Saclay, Batiment 709, 
91191 Gif sur Yvette, France \\[\affilskip]
$^3$Department of Astronomy, University of Washington, Seattle, WA, 98195, 
USA\\[\affilskip]
$^4$Carnegie Observatories, 813 Santa Barbara St., Pasadena, CA, 91101, USA}

\pubyear{2008}
\volume{254}  
\pagerange{666--671}
\jname{The Galaxy Disk in Cosmological Context}
\editors{J. Anderson, J. Bland-Hawthorn \& B. Nordstr\"om, eds.}
\begin{document}

\maketitle

\begin{abstract}
To date, fully cosmological hydrodynamic disk simulations to
redshift zero have only
been undertaken with particle-based codes, such as {\tt GADGET}, 
{\tt Gasoline}, or
{\tt GCD+}.  In light of the (supposed) limitations of traditional 
implementations of 
smoothed particle hydrodynamics (SPH), or at the very least, their
respective idiosyncrasies, it is important to explore 
complementary approaches to the SPH paradigm to galaxy formation.
We present the first high-resolution cosmological disk simulations 
to redshift zero using
an adaptive mesh refinement (AMR)-based hydrodynamical code, in 
this case, {\tt RAMSES}.  We analyse the temporal and spatial evolution
of the simulated stellar disks' vertical heating, velocity ellipsoids, 
stellar populations, vertical and radial abundance gradients
(gas and stars), assembly/infall histories, warps/lopsideness, 
disk edges/truncations (gas and stars), ISM physics implementations,
and compare and 
contrast these properties with our sample of cosmological 
SPH disks, generated with 
{\tt GCD+}.  These preliminary results are the first in our long-term 
Galactic Archaeology Simulation program.
\end{abstract}

\firstsection 
\section{Introduction}

The ability to form and evolve (correctly!) a disk galaxy with the aid of 
massively parallel computers and optimised algorithms remains an elusive
challenge for astrophysicists.  Ameliorating the non-physical effects 
associated with
overcooling, overmerging, angular momentum loss, and the capture
of accurate phenomenological prescriptions for the sub-grid physics
governing galaxy evolution (star formation, feedback, etc.) has been
achieved through rapid advancements in both hardware and software 
algorithms, but their complete elimination has yet to be realised.

Fully self-consistent cosmological hydrodynamic simulations of 
Milky Way-like disk galaxies, with sufficient resolution ($\simlt$500~pc)
to decompose and analyse various galactic sub-components (eg. halo, 
bulge, and thin + thick disks) have only really appeared over the 
past $\sim$5 years (Sommer-Larsen et~al. 2003;
Abadi et~al. 2003; Governato et~al. 2004,2007; Robertson et~al. 2004;
Bailin et~al. 2005; Okamoto et~al. 2005).  

A common thread
linking these studies is the use of a particle-based approach
to representing and solving the equations of hydrodynamics - usually
through the use of a smoothed particle hydrodynamics (SPH) code, such
as {\tt GADGET}, {\tt Gasoline}, or {\tt GCD+}.
Where there is no disputing the impact that SPH has had on the field, 
it is important to be aware of both the strengths {\it and}
weaknesses of any specific approach - as O'Shea
et~al. (2005) and Agertz et~al. (2007) have shown, both
subtle {\it and} overt differences can be introduced when employing
a particle-based, as opposed to a
mesh-based (or grid-based) approach (and \it vice versa\rm), when simulating
galaxy formation and evolution.

To address these concerns, we have initiated a long-term 
Galactic Archaeology Simulation programme aimed at complementing the
aforementioned particle-based studies (including our own)
with a comprehensive suite
of simulations generated with a grid-based 
N-body + hydrodynamical code employing
adaptive mesh refinement (AMR) -  our software tool of choice has been
{\tt RAMSES} (Teyssier 2002).  \it These simulations represent
(to our knowledge) the first to be generated through to
redshift zero, with a grid code, within
a fully cosmological and hydrodynamic framework.\rm\footnote{The
beautiful simulations of Ceverino \& Klypin (2008),
generated with the {\tt ART} grid code were not 
(again, to our knowledge) run to redshift zero.}

In this contribution, we provide a brief summary of the 
methodology adopted, and highlight several {\it preliminary}
results associated with our analyses of the simulations'
disk kinematics, chemistry, disk edges / truncations, and 
assembly / infall histories.

\vspace{-4mm}
\section{Methodology}

From a parent 20~$h^{-1}$~Mpc $\Lambda$CDM collisionless particle
simulation, several representative 
$\sim$5$-$8$\times$10$^{11}$~M$_\odot$ halos were 
identified for higher-resolution re-simulation with the
full baryonic physics capabilities of {\tt RAMSES}.  Unlike 
previous studies, we placed essentially no \it a priori \rm 
restrictions during the halo selection process - ie, 
we did {\it not} purposefully select isolated, median-spin halos, 
with particularly quiescent assembly histories, in an attempt
to bias the selection towards a ``Milky Way-like'' halo.

The parent dark matter simulation was re-centred on the halo of
interest with now three nested areas of different mass resolution.
Only the central 512$^3$ coarse grid was then refined, up to 7 
additional levels, with the full suite of baryonic physics
included (eg. star formation, blast-wave supernovae feedback
parametrisation, chemical enrichment, UV background, metal-dependent
cooling, etc.), resulting in a formal spatial (baryonic
mass) resolution of 435~pc (10$^6$~M$_\odot$) at $z$=0.\footnote{At
the time of writing, an
additional level of refinement has been completed, taking
the resolution to $\sim$200~pc.}  The resolution is roughly a factor
of two better than that employed in our earlier SPH work
(Brook et~al. 2004; Bailin et~al. 2005).

\vspace{-4mm}
\section{Basic Characteristics}

Our first
two $\sim$L$^\ast$ disks (Ramses1 and Ramses2, respectively) both
ended as fast-rotating massive (7.6$\times$10$^{11}$~M$_\odot$ and
5.5$\times$10$^{11}$~M$_\odot$, respectively) galaxies in low-spin
($\lambda$=0.02) halos.  Bandpass-dependent bulge-to-disk (B/D)
decompositions show, not surprisingly, B/D ratios in the range
of $\sim$0.4$-$0.8, signatures of the same overcooling / 
overcentralisation ``problems'' which plague traditional SPH disk
simulations.  The stellar bulge has a $V/\sigma$$\sim$0.5, reflecting
its $\sim$70~km/s rotation.\footnote{Similar to that of the Milky Way, although
we should emphasise again that Ramses1 is not supposed to be 
a one-to-one ``clone'' of the Milky Way.}
That said, the simulated I-band images (edge-on and 
face-on) for Ramses1 (middle and right panels of Fig~1) are 
more than encouraging.  The left-most panel of Fig~1 shows the 
gas density distribution (different projection); the tell-tale
warp can be traced to a neighbouring satellite.

\begin{figure}[htb]
\begin{center}
 \includegraphics[width=1.72in]{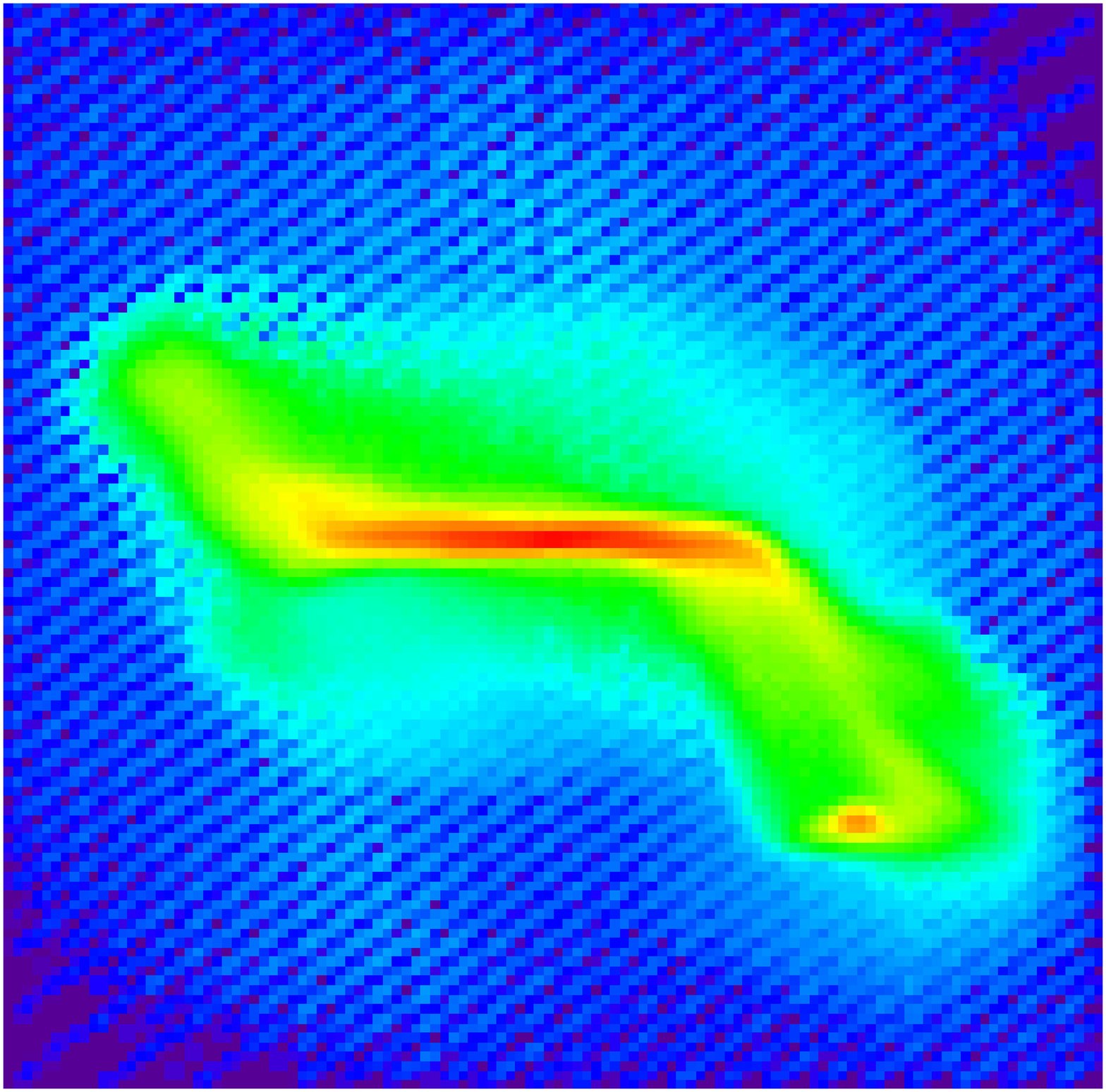}
 \includegraphics[width=1.70in]{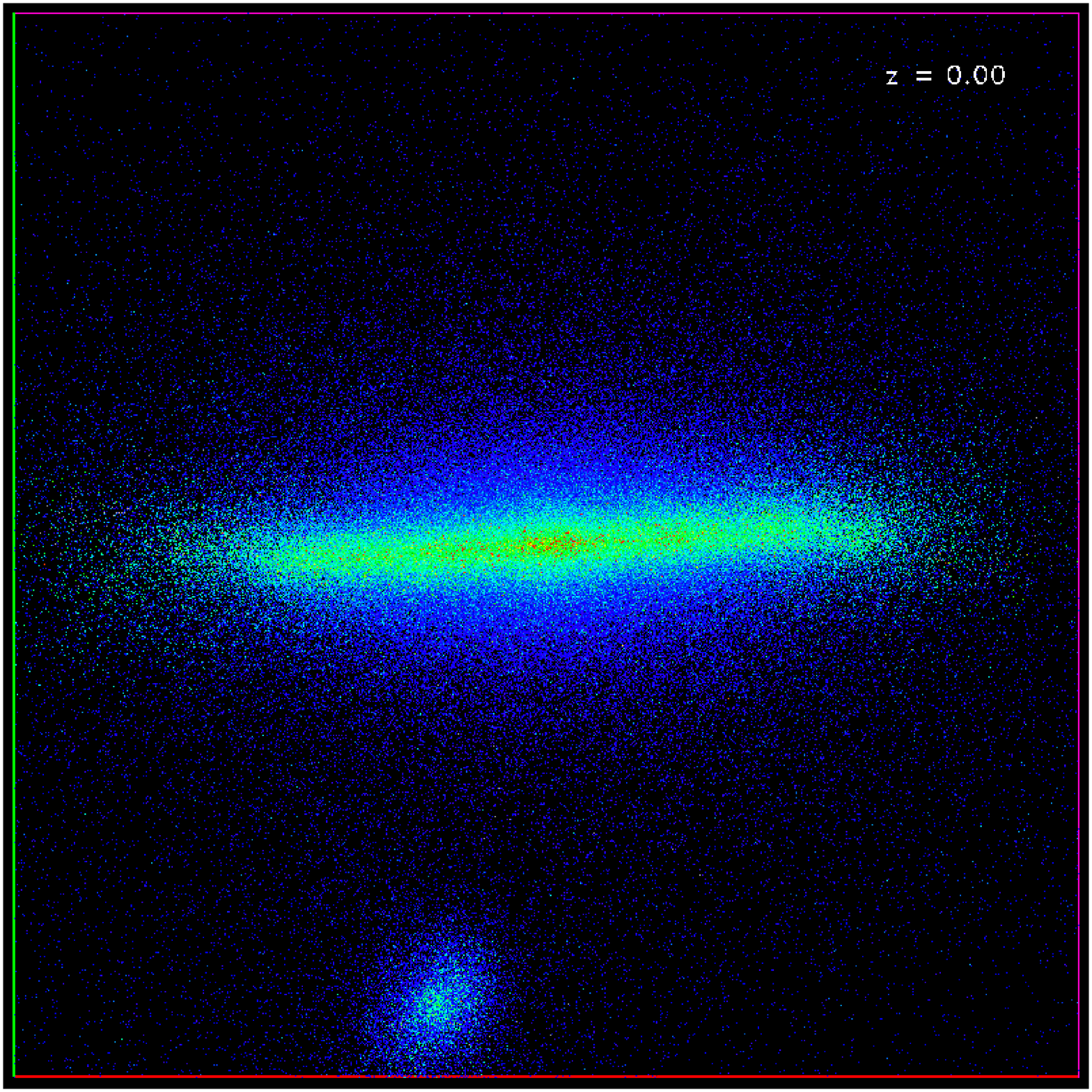}
 \includegraphics[width=1.70in]{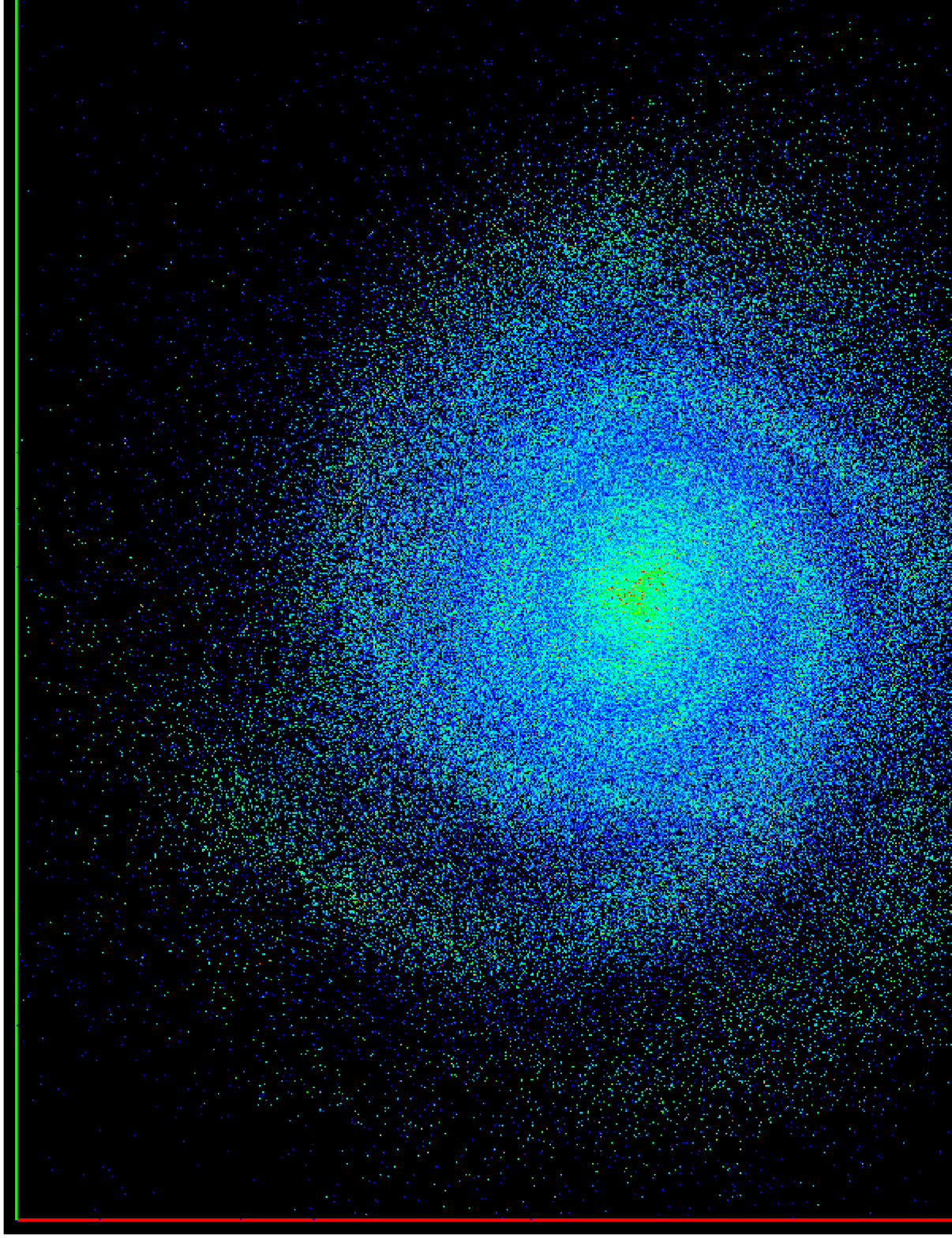}
 \caption{Gas (left) and stellar light (I-band: middle and right)
	distributions (60$\times$60~kpc) for the Ramses1 disk galaxy.}
   \label{fig1}
\end{center}
\end{figure}

The overall star formation histories for the two {\tt RAMSES} disks
are not dissimilar to those associated with our two
{\tt GCD+} SPH disks - GCD1 (Bailin et~al. 2005: fully-cosmological, using
the Abadi et~al. (2003) initial conditions; GCD2 (Brook et~al. 2004)
semi-cosmological (Fig~2).  Each shows the tell-tale peak in star formation
between $z$$\approx$2$-$4, with an associated exponential decline
over the past $\sim$10~Gyrs to a present-day rate of $\sim$1$-$2~M$_\odot$/yr.
In detail, the star formation histories of both {\tt RAMSES} disks
compare very favourably to that inferred 
from semi-numerical Galactic Chemical Evolution models, for the
Milky Way as a whole; indeed, Ramses2 and the Milky Way model
of Fenner et~al. (2005), are extremely similar in their global star formation
histories.

\begin{figure}[htb]
\begin{center}
 \includegraphics[width=3.40in]{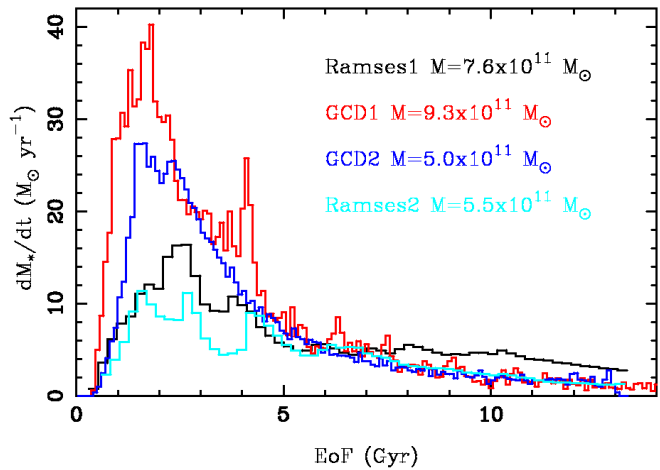}
 \caption{Time evolution of the star formation rate (within 30~kpc) for
	the two {\tt RAMSES} disks described here, alongside those for
	our two earlier {\tt GCD+} SPH disk galaxies.}
   \label{fig2}
\end{center}
\end{figure}

\vspace{-4mm}
\section{Chemistry}

Our current implementation of chemistry within {\tt RAMSES} is 
restricted to the global metal content (Z), under the assumption
of the instantaneous recycling approximation.  We are in the midst
of porting to {\tt RAMSES} the more sophisticated chemical evolution modules
from our {\tt GEtool} (95 isotopes; 34 elements: Fenner et~al. 2005)
and {\tt GCD+} (9 isotopes; 9 elements: Kawata \& Gibson 2003)
codes.
Having said that, there are useful and important chemical
characteristics which can be extracted and examined, including the 
degree of azimuthal variation in the global metallicity (both
stellar and gas-phase) and the overall metallicity gradients in the
thin and thick stellar disk components.

For example, within the Milky Way, the local peak-to-peak scatter in 
the gas-phase abundances is
$\sim$0.4~dex (Cescutti et~al. 2007; Fig~3).  Examining the Ramses1
simulation, we find a comparable azimuthal variation in
the gas metallicities ($\sim$0.2$\rightarrow$$-$0.5~Z$_\odot$) at
radii of 8$-$10~kpc.  We find a mid-plane (thin
disk) abundance gradient of dZ/dR$\sim$$-$0.03~dex/kpc, 
comparable to that observed locally (Cescutti et~al.; Tbl~5) and
consistent with an ``inside-out'' disk growth scenario (Fenner et~al.
2005); the gradient in the thick disk is a factor of two shallower over 
the same galactocentric distance.

\vspace{-4mm}
\section{Kinematics}

One of the pillars of Galactic Archaeology has been the 
suggestion that within the Milky Way, vertical disk heating saturates
at $\sigma_{\rm Z}$$\sim$20~km/s for stars of ages $\sim$2$-$10~Gyrs
(Quillen \& Garnett 2001); for older stars, 
a discrete jump to $\sigma_{\rm Z}$$\sim$45~km/s is apparent which
could be a signature
of the thick disk.
These conclusions have been questioned 
by Holmberg et~al. (2007), who claim that the evidence instead
supports a picture in which the
disk has undergone continual heating throughout
this period.  These opposing scenarios are represented in 
schematic form by the yellow lines in Fig~3.

Our semi-cosmological models (Brook et~al. 2004) appear to be
more consistent with
the ``disk saturation'' scenario (see the blue GCD2 curve of 
Fig~3), with the older, hotter, stars being associated with the
\it in situ \rm
formation of the thick disk during the intense gas-rich merger
phase at high-redshift.  The cosmological disks (both the new
{\tt RAMSES} pair, and our Bailin et al. 2005 SPH galaxy) though
appear to be more consistent (or at least not inconsistent) with the
``continual heating'' scenario (see the Ramses1, Ramses2, and GCD1
curves of Fig~3).  There are a large number of caveats that need
to be noted here, each of which will be explored in a future paper:
(i) the cosmological models capture late-time infall more accurately
than the semi-cosmological models; (ii) the Ramses2 model, while
showing evidence for continual disk heating, also shows evidence of
an ``impulsive'' step for stars older than $\sim$10~Gyrs, suggesting
a hybrid picture might be more appropriate for this galaxy; (iii) 
again, the cosmological disks were not chosen to be Milky Way 
``clones'', so one must be careful not to overinterpret the simulations.
Regardless, it is fascinating to see the four simulations filling the
area between the two extrema; we are expanding our simulation suite,
in order to explore the range of
heating scenarios, and its association with assembly history, environment,
and mass.

\begin{figure}[htb]
\begin{center}
 \includegraphics[width=3.40in]{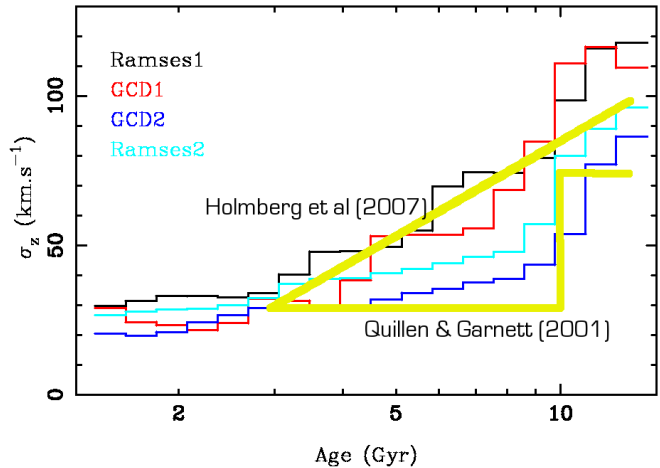}
 \caption{Age-velocity dispersion relation in the vertical 
	direction for ``local'' disk stars in our suite of 
	{\tt RAMSES} and {\tt GCD+} simulations.  The suggested
	behaviour in the solar neighbourhood under the ``disk 
	saturation'' (Quillen \& Garnett 2001) and ``continual 
	heating'' (Holmberg et~al. 2007) scenarios are shown in
	schematic form.}
   \label{fig3}
\end{center}
\end{figure}

We can go one step further and examine the spatial variation of 
$\sigma_{\rm Z}$ (over $\sim$1$-$4 disk scale lengths)
for intermediate-age stars (bottom right panel of Fig~5).  The
exponential decline observed is consistent with that
observed by Herrmann \& Ciardullo (2008) 
in six nearby face-on galaxies.  This is the
expected behaviour for disks with constant M/L; the relatively
high dispersions seen in the simulations 
(and in the
observational data) beyond $\sim$2$-$3 scale 
lengths is likely due to the combination of disk heating and 
flaring.

Finally, we note in passing that the thick disks in our
simulations lag those of their repsective thin disks, much
like the spirals in Yoachim \& Dalcanton (2008) (by $\sim$70$-$100~km/s
at galactocentric radii of 8$-$10~kpc and 3$-$6~kpc above the mid-plane)
This is encouraging, but
we next need to explore how this scales with mass (cf. the 
mass-dependent lag
claimed by Yoachim \& Dalcanton 2008), whether the thick disk 
scale lengths are consistently
greater than those of the thin disk (Yoachim \& Dalcanton 2006), 
and whether these are 
independent of environment, as claimed by Santiago \& Vale (2008).  Our
full suite of simulations will be brought to bear on these problems.

\vspace{-4mm}
\section{Disk Edges / Truncation}

\subsection{Gas Disks}

The Ramses1 disk has been simulated with a range
of ISM physics treatments, represented by models with
and without a polytropic equation of state.  Taking our simulation
without the polytrope, we examined the distribution of both the
neutral and ionised gas of the disk (see also 
Fig~1).  Ramses1 possesses a lopsided HI disk with a truncation/break
near 19~kpc, where the HI column density is
$\sim$2$\times$10$^{19}$~cm$^{-2}$ (red curve of Fig~4), consistent
with that observed by the THINGS team (Portas et~al. 2008). 
Admittedly, the break is not as clear as that observed empirically
(Portas et~al; Fig~3), but that reflects the fact that
we have azimuthally
averaged over 2$\pi$ radians, as opposed to splitting into twelve
$\pi$/6 segments and aligning at the ``break''; as such, the 
lopsidedness ``smears'' the break in Fig~4 from 19~kpc to a range
of radii spanning 19$-$26~kpc.  The ionised disk
extends $\sim$30$-$50\% beyond the neutral disk, before 
being ``lost'' in the background corona, similar to
that observed (Bland-Hawthorn et~al. 1997).

\begin{figure}[htb]
\begin{center}
 \includegraphics[width=3.40in]{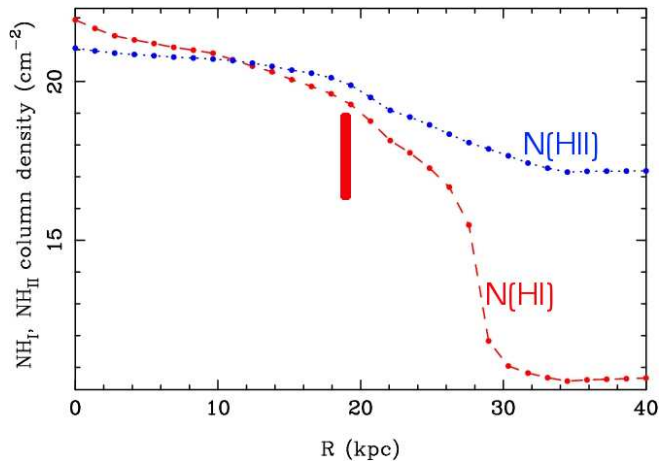}
 \caption{Neutral and ionised gas density distributions for the 
	``no polytrope'' simulation of Ramses1.  The edge of the
	HI disk occurs near 19~kpc (vertical line), 
	but is smeared out to $\sim$26~kpc
	by having averaged azimuthally over 2$\pi$ radians.}
   \label{fig4}
\end{center}
\end{figure}

\subsection{Stellar Disks}

The origin of the apparent truncations to the exponential disks 
seen in the surface brightness distributions
of spirals both locally and at high-$z$ is one of the
most exciting areas of disk galaxy ``astrophysics'' today (eg.
Bakos et~al. 2008 (B08); Roskar et~al. 2008 (R08); and many references therein).
The relative roles of star formation thresholds and radial 
migration / re-distribution of stars due to secular effects remains
hotly debated.

In Fig~5 (ignoring the bottom right panel, which refers
only to \S~5), we show a series of panels, based on our
analysis of the Ramses1 simulation, which should be 
examined beside the idealised simulation of
R08 (Fig~1) and the observational data of
B08 (Fig~1).  There are a number of similarities,
and tantalising differences, between the various datasets.

First, the two upper left panels show that while a break is seen 
at $\sim$10~kpc in both the B- and K-band, there is little (if any)
evidence for a break in the stellar surface density. 
In addition, the bottom middle panel shows the disk colour 
becomes blue with increasing radius prior to the break, but
becomes redder beyond it.  Each of these are in agreement with the
inferences derived empirically by B08; this is also more-or-less
in agreement with the conclusions derived by R08 from a non-cosmological
(but higher resolution) simulation, although R08 find a break in the 
stellar surface density that we (and B08) do not.  Much like the colour
becoming redder beyond the break, we find an associated increase in the 
stellar age of these stars (bottom left panel), 
in agreement with R08.  

\begin{figure}[htb]
\begin{center}
 \includegraphics[width=5.00in]{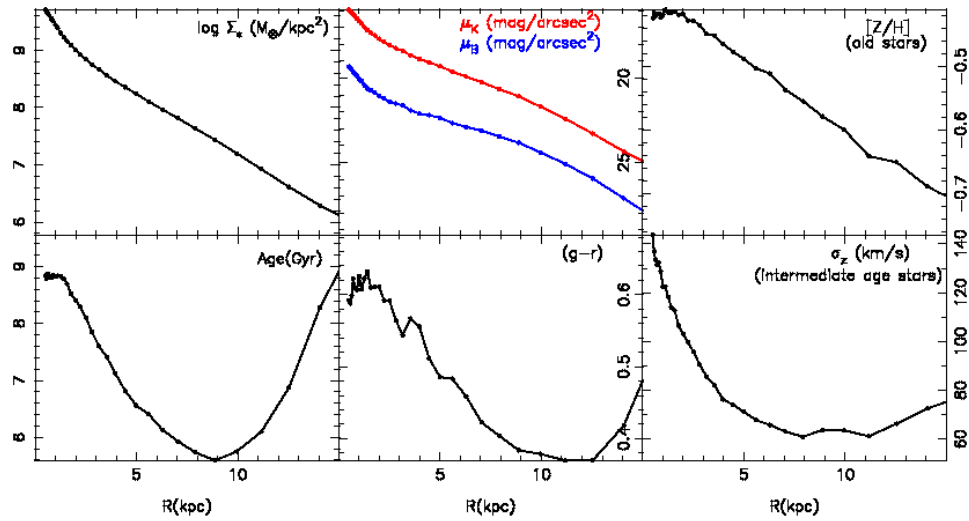}
 \caption{Radial stellar surface density (upper left), brightness 
(upper middle), mass-weighted metallicity for old stars
(upper right), mass-weighted age (bottom left), colour (middle bottom),
and vertical velocity dispersion (bottom right: \S~5) for the Ramses1
stellar disk.}
   \label{fig5}
\end{center}
\end{figure}

Within the picture proposed by R08, the stars
beyond the break today are primarily old, 
born primarily in the inner disk 
(somewhat inside the break radius at the time of birth), and migrated
to the outer disk due to various secular re-distribution effects.
We are exploring the veracity of this elegant suggestion with our
suite of cosmological simulations; 
there is migration within the cosmological simulation, but the 
relative contribution of secular re-distribution, \it in situ \rm
formation,\footnote{Which does occur, as there is a not
insignificant number of gas cells in the disk with densities in 
excess of the 0.1~cm$^{-3}$ star formation threshold.} and satellite
debris,\footnote{$\sim$50\% of stars beyond the break formed since $z$=2.3, 
the highest redshift for which we could accurately identify and
align the disk; of this fraction, $\sim$30\% formed at galactocentric
radii in excess of the present-day break; indeed, much of this
disk ``debris'' formed in satellites distributed fairly uniformly
in radius within the virial radius of the host halo.}
beyond the break, needs careful examination.

One issue which needs to be 
addressed, which has perhaps not been appreciated previously, 
is that shown in the upper right panel to Fig~5.  Here, we
show the metallicity gradient of the old ($>$7~Gyrs) stars in
our Ramses1 simulation.  From the bottom left panel, we see that
the outer disk stars are old; from the upper right panel, we see
that they are also relatively metal-poor; in and of itself, this
seems consistent with R08.  The potential problem which arises is that
from the upper right panel, we also see that stars of the same age in 
the inner disk are a factor of two more metal-rich.  \it One must ask
why it is that old metal-poor stars from the inner disk get
re-distributed to the outer disk, but old metal-rich stars of the 
same age from the inner disk do not get re-distributed. \rm 
We need to examine the metallicity distribution functions as a function
of space and time within the simulations (both our's and those of R08)
to better understand the situation.  

\vspace{-4mm}
\section{Gas Accretion / Infall}

We defer a detailed discussion of the gas accretion history
to the future, but felt it worth noting here that we have measured
the spatial and temporal infall of cold, warm, and hot gas both
into the halo as a whole and onto the disk itself.  For Ramses1, 
the flux of gas across a sphere of radius 30~kpc since 
redshift $z$$\sim$2 is $\sim$1.5$-$0.5~M$_\odot$/yr; this 
can be contrasted with the inferred infall within semi-numerical
Galactic Chemical Evolution models: for example, for our
Milky Way model described in Fenner et~al. (2005), averaging over the
entire disk, for the same redshift range, results in a predicted
infall rate of $\sim$5$-$1~M$_\odot$/yr.  We can also determine 
the metallicity distribution of this infalling gas; for this
particular 30~kpc surface at $z$=0, $<$Z$_{\rm g}$$>$$\sim$0.02~Z$_\odot$, 
with essentially no component in excess of $\sim$0.05~Z$_\odot$, 
again consistent with the Fenner et~al. semi-numerical models
(which assume the infalling gas is $\simlt$0.1~Z$_\odot$).
We have also laid virtual
slabs $\pm$6~kpc above and below the mid-plane of the simulated disk
and measured the vertical gas flux through these surfaces, finding 
a rate of $\sim$1~M$_\odot$/yr.  We have conducted the same
experiment with virtual slabs at $\pm$1~kpc above/below the plane, and
found fluxes $\sim$10$-$50~$\times$ greater, reflecting the 
much greater ``ISM circulation flux'' near the plane dominating over the
flux from cosmological infall, consistent with observations.

\vspace{-4mm}
\section{Summary and Future Directions}

We have realised (to the best of our knowledge)
the first fully self-consistent cosmological hydrodynamic
disk simulations to $z$=0 with a mesh code; the resolution attained
is 435~pc.  Several preliminary results include:
\begin{itemize}
\item the saturated vertical disk heating seen in semi-cosmological
SPH simulations has not yet been clearly replicated in our cosmological
simulations;
\item the neutral gas disks show ``edges'' at comparable column
densities to those observed; ionised gas disks extend beyond the
neutral gas, again in agreement with those observed;
\item the stellar surface brightness profiles show ``breaks'' in the 
exponential profiles, with associated increases (reddening) in the
age (colour) of the stellar populations beyond the break, in
agreement with observation; little evidence is seen for an associated
break in the stellar surface density profile, also as inferred from
observations; stars of the same age beyond and interior to the break
do not appear to have the same metallicity, which may prove problematic for
radial migration scenarios;
\item gas accretion is not smooth, but does
appear to be more-of-less ``inside-out'';
\item the disk-halo ``circulation flux'' is $\sim$10-50$\times$ that
of the ``infalling flux'' (again, consistent with the broad numbers
associated with the Milky Way).
\end{itemize}

Beyond the analysis of the extant simulations, 
we have a number of planned enhancements,
including full chemical evolution / tagging,
a ten-fold increase in the number of simulations (to examine
scaling relations, environmental dependencies, and assembly
history variations), a range of ISM physics implementations
(various polytropic equations of state, blast wave parametrisations),
quantifying warp and lopsidedness statistics (Mapelli et~al. 2008), 
2d IFU Lick index-style
maps, dusty radiative transfer, high-velocity clouds, radial gas
flows, and detailed SPH vs AMR comparisons with identical initial
conditions.

\vspace{-4mm}
\section*{Acknowledgements}
The support of the UK's Science \& Technology
Facilities Council (ST/F002432/1) and the Commonwealth 
Cosmology Initiative are gratefully acknowledged.  
We also wish to thank Ignacio Trujillo and Judit Bakos for their
guidance.  Simulations and
analyses were carried out on COSMOS (the UK's National Cosmology 
Supercomputer) and the University of Central Lancashire's
High Performance Computing Facility.  The parent N-body simulation was 
performed within the framework of
the Horizon collaboration ({\tt http://www.projet-horizon.fr).

\end{document}